\documentclass[12pt]{iopart}
\usepackage{graphicx}
\def\half{\frac{1}{2}}

\def\beq{\begin{eqnarray}}
\def\eeq{\end{eqnarray}}

\begin{document}
\title{A non-perturbative method for time-dependent problems in quantum mechanics}
\author{Paolo Amore\footnote{paolo@ucol.mx}
Alfredo Aranda\footnote{fefo@ucol.mx}}
\address{Facultad de Ciencias, Universidad de Colima, \\
Bernal D\'{\i}az del Castillo 340, Colima, Colima, M\'exico}

\author{Francisco M. Fern\'andez \footnote{fernande@quimica.unlp.edu.ar}}
\address{INIFTA (Conicet,UNLP), Diag. 113 y 64 S/N, \\
Sucursal 4, Casilla de Correo 16, 1900 La Plata, Argentina}

\author{Hugh Jones \footnote{h.f.jones@imperial.ac.uk}}
\address{Department of Physics, Imperial College, \\
London SW7 2AZ, England}

\begin{abstract}
A powerful method for calculating the  eigenvalues of a Hamiltonian operator
consists of converting the energy eigenvalue equation into
a matrix equation by means of an appropriate basis set of functions.
The convergence of the method can be greatly improved by means of a variational
parameter in the basis functions determined by the principle of minimal sensitivity.
In the case of the quartic anharmonic oscillator and of a symmetrical double-well potential
we choose an effective oscillator frequency. In the case of nonsymmetrical potential we add a coordinate shift in a
two-parameter variational calculation.
The method not only gives the spectrum, but
also an approximation to the energy eigenfunctions. Consequently it can
be used to solve the time-dependent Schr\"odinger equation using the method
of stationary states. We apply it to the time development of two different
initial wave functions in the double-well slow roll potential.
\end{abstract}
\maketitle

\section{Introduction}

We present a method for obtaining arbitrarily
precise approximations to the solution of the time-dependent
Schr\"odinger equation with a potential which fulfills the
condition $\lim_{|x|\rightarrow \infty} V(x) = + \infty$ (i.e. a
potential which only admits bound states). Although there are many
examples of potentials of this kind, only a limited number of them
can be solved exactly, the best known example being the simple
harmonic oscillator (SHO), which is a standard topic in virtually
any quantum mechanics textbook and which can be used to model many
physical systems.

In this paper we consider problems which cannot be solved
analytically and where an alternative strategy must be found.
Perturbation theory  is the standard tool which is used to
deal with such problems; unfortunately, the straightforward application
of perturbation theory to some problems is not practical because the
perturbation series is divergent. There are various methods to overcome this
apparent limitation; for example the linear delta expansion (LDE)~\cite{DM, HFJ} and
other variants of variational perturbation theory
(VPT)~\cite{AFC90, Jan95, Yuk91, AAD:04,AADL:04}. Loosely
speaking, these techniques, although differing in the details, are
based on the powerful idea that one can obtain a new expansion in
some ``unnatural'' parameter (i.e. one not appearing in the original
problem) and that the sequence of approximants resulting from this expansion can be
made to converge very fast by suitably choosing a variational
parameter. For example, the linear delta expansion works by
interpolating the full Hamiltonian with the Hamiltonian
corresponding to a soluble model, which depends on an arbitrary
parameter, and by applying perturbation theory to it. The
parameter is then determined by means of the principle of minimal
sensitivity (PMS)~\cite{PMS}. Since the optimum value of the adjustable parameter given by the
PMS depends upon the natural parameters in the Hamiltonian, the result
corresponds to a non-perturbative result, i.e. to a non-polynomial
expression in the natural parameters.

Among other applications of the LDE and VPT we mention an improved
Lindstedt-Poincar\'e method~\cite{AA1:03, AA2:03, AM:04}, the
calculation of the period of classical
oscillators~\cite{AS:04,AAFS:04,AF:04}, the spectrum of a quantum
potential with the WKB method~\cite{AL:04} and the acceleration of
the convergence of mathematical series~\cite{Amore:04}.
However, it was found that the LDE fails to give the correct long-time behavior of the
 wave-function in the quantum mechanical version of the slow-roll
scenario of inflation\cite{Jones:2000au}.
In successive orders it is able to approximate
the exact time development more and more accurately, but only up to the
time where the wave function has spread out and is beginning to contract again.
The Hartree-Fock method does give a general qualitative picture of the time-development
at later times, but is very far from being accurate.

Here we propose an alternative method that consists of converting the eigenvalue equation into a
matrix equation by taking matrix elements with respect to harmonic
oscillator wave functions of arbitrary frequency $\Omega$, and then determining
$\Omega$ by some version of the PMS criterion. The particular criterion used
here was first proposed and utilized in \cite{QM85}. Having thus obtained
the approximate energy eigenvalues and eigenfunctions, one can then use the
method of stationary states to calculate the time-dependence of the state for a given initial
configuration. It turns out that this method is extremely accurate at even
quite small orders, and has no problem with the long-time behaviour.

The article is organized as follows: in Section \ref{spectrum} we describe the
method in general terms and apply it to calculating the spectrum of various anharmonic oscillators;
in Section \ref{time-dependence} we use the method
of stationary states to follow the time development of two initial wave-functions
for the double-well potential that has been used in slow roll inflation and
compare our results with  those found in the literature;
finally, in Section \ref{concl} we draw our conclusions.

\section{Energy Spectrum}\label{spectrum}
\subsection{The Method}

We tackle the problem of solving the energy eigenvalue equation
\begin{eqnarray}
\hat{H} \psi_n = E_n \psi_n \label{eq2}
\end{eqnarray}
by converting it to a matrix equation, using an orthonormal basis
of wave functions of the quantum harmonic oscillator, depending upon an arbitrary frequency
$\Omega\equiv \alpha^2$:
\begin{eqnarray}
\phi_n(x) = N_n \ e^{- \alpha^2 x^2/2} \ H_n(\alpha x) \ ,
\label{eq4}
\end{eqnarray}
where the normalization constant is $N_n =(\alpha/(2^n \ n! \ \sqrt{\pi}))^\half$.
$\hat{H}$ is the Hamiltonian for a particle in a one-dimensional
potential that supports only bound states.

It is necessary to truncate the infinite-dimensional matrix $H_{n\ell}$ to some
finite dimension, say $N\times N$, and then its
eigenvalues can be calculated by simple matrix diagonalization.
It is known that as $N$ increases the approximation for the
energy levels should steadily improve. This is indeed
the case, but for an arbitrary $\Omega$ the convergence may be
quite slow. This leads one to seek for some criterion to choose an
optimum value of $\Omega$. The criterion we shall adopt here, which
is essentially that adopted in \cite{QM85}, is the principle of
minimal sensitivity\cite{PMS} applied to the trace of the truncated
matrix.

The rationale behind this principle is that the eigenvalues and other exact quantities
of the problem are independent of $\Omega$ but any approximate result coming from the
diagonalization method for finite $N$ exhibits a spurious dependence on the
oscillator frequency. This also applies to the trace
of the matrix, the sum of the eigenvalues. However, for finite $N$ a
spurious $\Omega$ dependence will emerge in the sum. A reasonable criterion for
choosing $\Omega$ is therefore to take it at a stationary point of
${\cal T}_N\equiv \sum_{n=0}^{N-1} H_{nn}$, so that this invariance is
respected, at least locally. Thus we impose the PMS condition
\begin{eqnarray}
\frac{\partial}{\partial \Omega} {\cal T}_N = 0 \ .
\label{pms}
\end{eqnarray}
The reason for applying this condition to the trace is that ${\cal T}_N$
is a simple quantity to evaluate, and moreover it is invariant under the unitary transformation
associated with a change of basis. Once $\Omega$ is so determined,
one obtains an approximation to the first $N$ eigenvalues
and eigenvectors of $\hat{H}$ by a numerical diagonalization of the truncated
$N \times N$ matrix. One could also contemplate applying the PMS
to the determinant, which of course shares the property of invariance
under unitary transformations, but this would be a much more cumbersome
calculation, and could well introduce many spurious PMS values.

In the following sections we will need the harmonic oscillator
matrix elements of $x^p$. Closed formulas have been given in
Ref.~\cite{Morales}, which we adapt here for completeness.
\beq
(x^{2r})_{n\ell}=\frac{\sqrt{n!\ell!}}{(2\sqrt{\alpha})^{2r}}\sum_{k=0}^{{\rm min}(n,r-\lambda)}\frac{(2r)!}
{2^{2r-k-\lambda}(r-\lambda-k)!(n-k)!(2\lambda+k)!k!}
\eeq
for $\ell-n=2\lambda$, and
\beq
\hspace{-2cm}(x^{2r+1})_{n\ell}=\frac{\sqrt{n!\ell!}}{(2\sqrt{\alpha})^{2r+1}}\sum_{k=0}^{{\rm min}(n,r-\lambda)}\frac{(2r+1)!}
{2^{2r-k-\lambda+\half}(r-\lambda-k)!(n-k)!(2\lambda+1+k)!k!}
\eeq
for $\ell-n=2\lambda+1$. These formulas assume $\ell\ge n$, but the matrix
is symmetric. In both cases $r$ must be greater than $\lambda$. For all other
values of $p$, $n$ and $\ell$ the matrix elements vanish.

In addition to these we will need the matrix elements of $p^2$, which are
given by
\beq
\hspace{-1cm}(p^2)_{n \ell} = -\frac{1}{2}\alpha^2\left[\sqrt{\ell(\ell-1)}\delta_{\ell,n+2}-(2\ell+1)\delta_{\ell,n}
+\sqrt{(\ell+1)(\ell+2)}\delta_{\ell,n-2}\right]
\eeq

\subsection{The quartic anharmonic oscillator}
We take the Hamiltonian as
\beq
\hat{H}=\half (p^2+ m^2 x^2)+ gx^4,
\label{eq7}
\eeq
whose matrix elements in the basis of the wave functions of Eq.({\ref{eq4}) are
\begin{eqnarray}\label{matrixelem}
H_{n\ell} &=&  \half (p^2)_{n\ell}+ \half m^2 (x^2)_{n\ell}+ g(x^4)_{n\ell}.
\end{eqnarray}
The trace to order $N$ is given by
\beq
\frac{4}{N}{\cal T}_N = N(\Omega-\frac{m^2}{\Omega})+\frac{g}{\Omega^2}( 1+2N^2).
\eeq
It turns out that ${\cal T}_N$ has a single minimum, located at
\beq
\Omega_{PMS} = \frac{m^2}{X^{1/3}}-\frac{1}{3}X^{1/3},
\eeq
where
\beq
X \equiv \frac{3}{N} \left[ -9g( 1+2N^2)
+ {\sqrt{3}} \left(-N^2m^6+27g^2(1+2N^2)^2\right)^\frac{1}{2}\right]
\eeq

\begin{figure}
\begin{center}
\includegraphics[width=9cm]{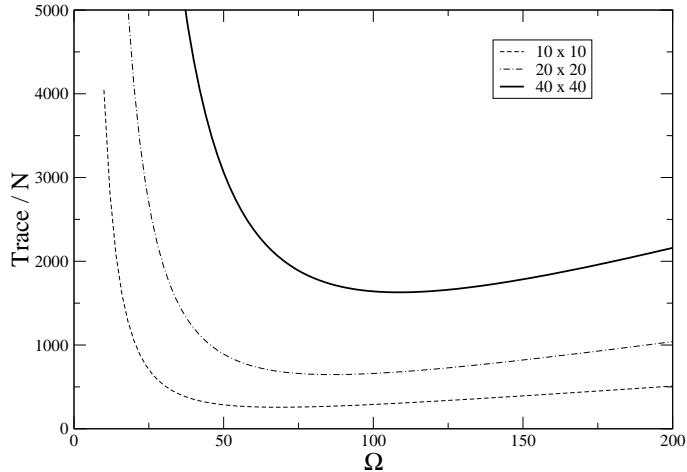}
\caption{Trace of the hamiltonian matrix~(\ref{matrixelem})
normalized by the number of states as a function of the
variational parameter $\Omega$.
We use the parameters  $m = 1$, $g = 4000$, corresponding to the case
studied in \cite{mei97}. Different curves represent different subspaces.}
\label{FIG2a}
\end{center}
\end{figure}

The graph of this trace (divided by $N$) against $\Omega$ is shown in Fig.~\ref{FIG2a}, exhibiting
a global minimum. Here we have taken the parameters as $m=1$, $g=1000$, and later multiplied the eigenvalues
by a factor of $2$ for comparison with the results of \cite{mei97} corresponding to $\beta=2000$.
By taking $\Omega$ at the minimum, in accordance with PMS, we obtain a precise approximation to the spectrum.

The error in the energy of the $n${th} excited state, i.e.
$\Delta \equiv |E_n^{(N)}-E_n^{(100)}|$, decreases exponentially
with $N$, as is shown in Fig.~\ref{FIG1} for the ground state 
and for the $49$th excited state.  $E_n^{(N)}$ is the energy of the $n$th
state obtained by considering the $N \times N$ subspace, whereas
$E_n^{(100)}$ is the energy of the $n$th state corresponding to
the largest subspace considered in this paper, which is used as
reference.

For $N=100$ we obtain the energy of the ground state of the Hamiltonian $p^2+x^2+2gx^4$ as
$$E_0^{(100)} = \underline{13.38844170100806193900617690280728652296098988517435666039}9$$
which agrees in the first $58$ (underlined) digits with the corresponding result of Table ~1 of \cite{mei97},
obtained with a different method. Much higher precision can be obtained by enlarging the $N \times N$ subspace:
this costs little additional effort because, once the PMS has been applied, the Hamiltonian matrix is fully numerical and
its eigenvalues/eigenvectors can be calculated numerically with accuracy and speed.

It is well known that the accuracy of the eigenvalues given by the method of Rayleigh-Ritz decreases as the quantum number increases:
this happens because the influence of the states which lie outside the subspace $N \times N$ is felt more strongly by the ``border states'',
i.e. those states which fall on the border of the selected subspace.
In order to calculate highly excited states without enlarging the subspace too much one can center the subspace 
around that particular state and apply the procedure mentioned above.
This has been done in Fig.~\ref{FIG1a}, where we have plotted the ratio
$\Delta_n = (E_n^{(N)}- E_n^{(WKB)})/E_n^{(WKB)}$, considering
a square subspace of elements $H_{ij}$ centred around the $n=100$ state.
$E_n^{(WKB)}$ is obtained using the analytical formula for the energy of the quantum anharmonic oscillator obtained in~\cite{AL:04}.
The flattening of the error around $n$ for the largest subspaces considered ($81 \times 81$ and $161 \times 161$) signals that
the numerical results obtained with the present method have reached the precision of the WKB formula.
We conclude that the present method can be used to obtain the energies and wave functions of arbitrarily high
excited states with the desired accuracy.
As expected, the error $\Delta_n$ is maximal for the ``border states'',

When the centred subspace is restricted to one element $N=1$ the
method reduces to a well--known simple variational calculation.

\begin{figure}
\begin{center}
\includegraphics[width=9cm]{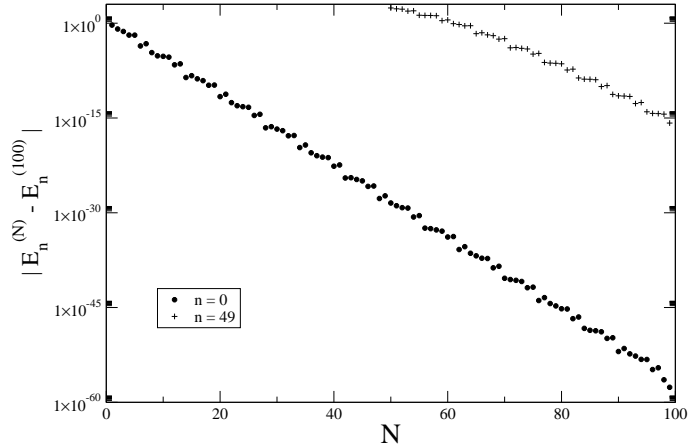}
\caption{Error in the energy of the ground and $49^{th}$ excited states as a function of $n$ and for subspaces of different dimensions.
We use the parameters  $m = 1$, $g = 4000$, corresponding to the case
studied in \cite{mei97}.}
\label{FIG1}
\end{center}
\end{figure}

\begin{figure}
\begin{center}
\includegraphics[width=9cm]{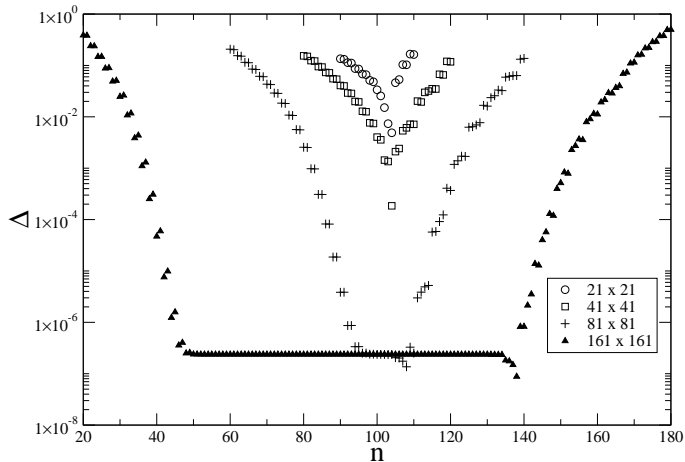}
\caption{$\Delta_n =(E_n^{(N)}-E_n^{(WKB)})/E_n^{(WKB)}$ as a function of the dimension of the subspace.}
\label{FIG1a}
\end{center}
\end{figure}

\subsection{The double-well potential}

The same method can be applied with very little change to the Hamiltonian
which has been used in the consideration of slow roll inflation in
the early universe, namely
\begin{eqnarray}
H&=&\half p^2 +
\lambda(x^2-a^2)^2/24 + const.\cr
&&\cr
&=& \half p^2 -\half m^2 x^2 +
gx^4,
\label{dwell}
\end{eqnarray}
with  $m^2=\lambda a^2/6$ and $g=\lambda/24$.

The parameters will be taken as $a = 5$ and $\lambda =  0.01$, as these have been used in previous
work on the subject (see, for example,\cite{Jones:2000au,Guth:1985ya,Cooper:1986wv})

All that is needed in this case is to reverse the sign of $m^2$ in
the formulas given in the previous subsection. We obtain similar
accuracy for the eigenvalues with very little effort. These will
be used in Section 3 to give the time development of a given
initial configuration.

\subsection{General anharmonic potentials}

We now consider general anharmonic potentials of the form:
\begin{eqnarray}
V(x) = \sum_{j=0}^{\cal N} \kappa_j x^j
\end{eqnarray}
where the coefficients $\kappa_j$ define a polynomial of order  ${\cal N}$ (even).
We require that $\kappa_{\cal N}>0$ to ensure that only bound states are permitted.

The shifted potential
\begin{eqnarray}
V(x+\sigma) = \sum_{j=0}^{\cal N} \kappa_j \ (x+\sigma)^j=\sum_{j=0}^{\cal N} \kappa_j \ \sum_{k=0}^j {j\choose k} x^{j-k}\sigma^k \ ,
\end{eqnarray}
and the original one have the same spectrum. Therefore we can choose $\sigma$ to be a variational parameter and apply the PMS
as in the case of the adjustable oscillator frequency.

As an example, let us consider the potential:
\begin{eqnarray}\label{Vasym}
V(x) = 11 - 118  x - 44 x^2 + 80 x^3 + 16 x^4 \ .
\end{eqnarray}
Since $V(x)$ is strongly asymmetric we expect that decomposing it with respect to a basis of SHO wave functions
centred at the origin will not be the best choice. We therefore translate the potential\cite{Maluendes1982} by an arbitrary
quantity $-\sigma$
and impose  the PMS on $\Omega$ and $\sigma$ simultaneously. That is, we impose the
two PMS conditions
\begin{eqnarray}
\frac{\partial {\cal T}_N }{\partial \Omega}   = 0  \ \ \ , \ \ \
\frac{\partial {\cal T}_N }{\partial \sigma}   = 0 \ .
\label{pms2}
\end{eqnarray}

Using a $10 \times 10$ subspace we obtain the optimal values
$\sigma = -3.889$ and $\Omega = 31.179$. Correspondingly we find
the energy of the ground state to be:
\begin{eqnarray}
E_0^{(10)} = -\underline{1229.11605104}5
\end{eqnarray}
where the underlined digits are correct. Using a $40 \times 40$
subspace we find $\sigma = -3.583$ and $\Omega = 27.431$,
obtaining
\begin{eqnarray}
E_0^{(40)} = -\underline{1229.11605104600459705899}2
\end{eqnarray}
where the underlined digits are correct. Similar results hold for the excited states.

We notice that $\sigma_{PMS}$ does not correspond to the value for which $V(x)$ has a minimum,
i.e. $x \approx - 3.979$. In fact $\sigma_{PMS}$ tends to increase as the dimension of the subspace is increased, as a result of the influence of the highly excited
states, which are less localized.

It is clear that such a scheme can be applied with limited effort to a general anharmonic potential of arbitrary order: in fact,
a suitable choice of $\sigma$ allows one to reach high  precision with a limited number of terms. Clearly, since a
general potential can be always expanded in a Taylor series around a point, this implies that our method can be easily
applied to calculate portions of the spectrum of a potential with arbitrary precision.

In Fig.~\ref{asym} we have plotted the first $200$ states of the spectrum of Eq.~(\ref{Vasym}),
using  our method with subspaces of $100 \times 100$ and $200 \times 200$ and  $500 \times 500$, together with
the first-order WKB estimate for comparison.

\begin{figure}[ht]
\begin{center}
\rotatebox{0}{\includegraphics[width=9cm]{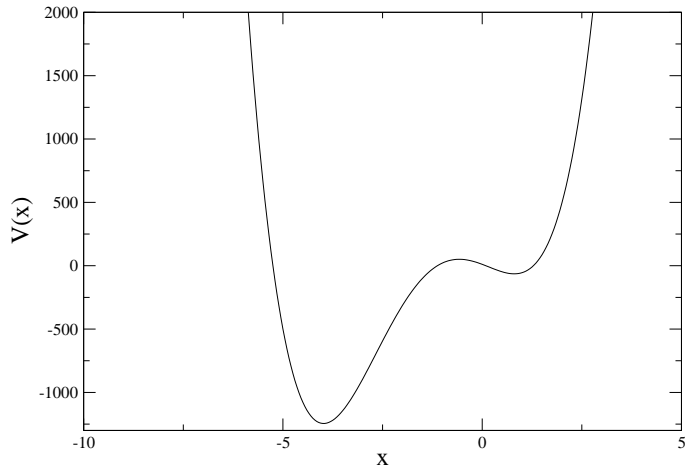}}
\end{center}
\caption{Potential $V(x) = 11 - 118  x - 44 x^2 + 80 x^3 + 16 x^4$.}
\label{potential}
\end{figure}

\begin{figure}[ht]
\begin{center}
\rotatebox{0}{\includegraphics[width=14cm]{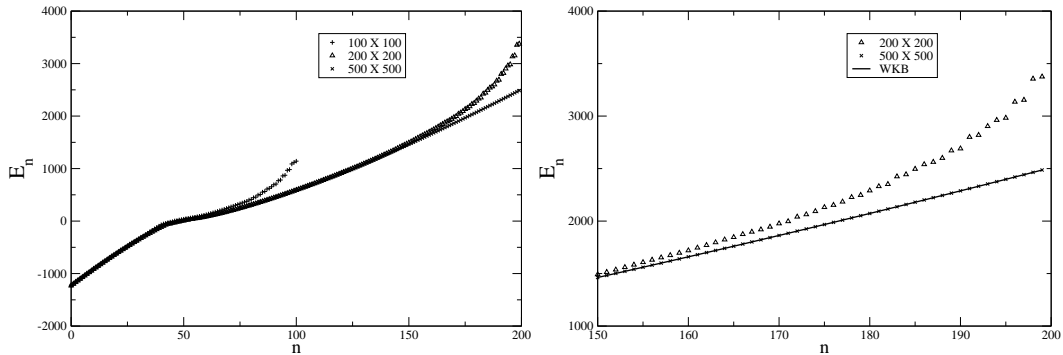}}
\end{center}
\caption{Spectrum of the potential of Eq.~(\ref{Vasym}), as a function of the quantum number $n$. The
pluses, triangles and crosses correspond to the eigenvalues obtained
using our method with subspaces of $100 \times 100$, $200 \times 200$ and $500 \times 500$ respectively. The
first-order WKB estimate is also shown for the higher states.}
\label{asym}
\end{figure}

\section{Time development}
\label{time-dependence}

\subsection{The Method of Stationary States}
If solutions of the energy eigenvalue equation (\ref{eq2})
are known for a given Hamiltonian $\hat{H}$ then the time-dependent
Schr\"odinger equation ($\hbar=1$)
\begin{eqnarray}
i \frac{d}{dt} \Psi = \hat{H} \Psi \  . \label{eq1}
\end{eqnarray}
can be solved by the method of stationary states. Namely, if the
initial wave-function at $t=0$ can be expanded as
\begin{eqnarray}
\Psi(x, t=0) = \sum_{n=0}^\infty a_n \psi_n(x) \label{InitWF},
\end{eqnarray}
its value at a later time is given  by
\begin{eqnarray}
\Psi(x,t) = \sum_{n=0}^\infty a_n \ e^{-i E_n t} \ \psi_n(x)
\label{eq3}
\end{eqnarray}

The method of Section \ref{spectrum} gives us an approximation not only to the spectrum, but
also to the energy eigenfunctions, namely
\begin{eqnarray}
\psi_n(x) = \sum_{k=0}^{N-1} {d}_{nk}  \ \phi_k(x) \ .
\label{eq5}
\end{eqnarray}
where $d_{nk}$ denotes the $k$th component of the $n$th eigenvector of the  truncated Hamiltonian matrix.

Similarly, the initial wave function can also be expressed as a truncated expansion in terms of the $\phi_n(x)$:
\begin{eqnarray}
\Psi_n(x, t=0) = \sum_{k=0}^{N-1} c_n  \ \phi_n(x) \ .
\label{eq55}
\end{eqnarray}

By comparison with Eq.~(\ref{eq5}) we see that
\begin{eqnarray}
a_n = \sum_{\ell=0}^{N-1} c_\ell \ ({\bf d}^{-1})_{\ell n} \ .
\label{eq555}
\end{eqnarray}

These coefficients can now be used in Eq.~(\ref{eq3}) to obtain an approximation to the wave function
$\Psi(x,t)$ at any later time $t$.

\subsection{Slow roll inflation}

Here we use the Hamiltonian of Eq.~(\ref{dwell}) with two different initial configurations.

\subsubsection{Centred Gaussian}\hfill\\

The initial wave function, used in previous studies of slow-roll inflation, is given by
$$
\Psi(x,t=0) = \left(\frac{m}{2\pi} \right)^{1/4}
e^{- m x^2/4}.
$$
Our task is to find the coefficients $c_n$ in Eq.~(\ref{eq55}).

By orthonormality,
\begin{eqnarray}
c_n = \int \phi_n^*(x) \Psi(x,t=0) dx = N_n \ \left(\frac{m}{2\pi} \right)^{1/4} \
\int_{-\infty}^{+\infty}  e^{- \beta^2 x^2/2} H_n(\alpha x) dx
\end{eqnarray}
where $\beta^2 \equiv m/2 + \alpha^2$. By means of the
change of variable $y = \alpha x$
\begin{eqnarray}
c_n = \frac{N_n}{\alpha} \left(\frac{m}{2\pi} \right)^{1/4} \int_{-\infty}^{+\infty}
e^{- \frac{\beta^2}{2 \alpha^2} y^2} H_n(y) dy \ .
\end{eqnarray}
and the expansion of $H_n(y)$, namely
\begin{eqnarray}
H_n(y) &=& \sum_{k=0}^{[n/2]} (-1)^k \frac{n!}{(n-2 k)! k!} 2^{n-2 k} y^{n -2 k} \ .
\end{eqnarray}
we obtain 
\begin{eqnarray}
c_n &=& \frac{N_n}{\alpha} \left(\frac{m}{2\pi} \right)^{1/4}
\sum_{k=0}^{[n/2]} (-1)^k \frac{n!}{(n-2 k)! k!} 2^{n-2 k} \ J_k
\end{eqnarray}
where
\begin{eqnarray}
J_k &\equiv& \int_{-\infty}^{+\infty}   y^{n -2 k}  e^{- \frac{\beta^2}{2 \alpha^2} y^2} dy
= \left( \frac{\sqrt{2} \alpha}{\beta} \right)^{n-2 k+1} \ \int_{-\infty}^{+\infty} z^{n-2 k} \ e^{-z^2} dz \ . \nonumber
\end{eqnarray}

In fact $c_n = 0$ unless $n$ is even $n = 2 \ell$, and then
\begin{eqnarray}
\hspace{-2cm}c_{2 \ell} &=& \frac{N_{2 \ell}}{\alpha}  \left(\frac{m}{2\pi} \right)^{1/4}
\sum_{k=0}^{\ell} (-1)^k \frac{(2 \ell)!}{(2 \ell-2 k)! k!} 2^{2 (\ell -k)}
\left( \frac{\sqrt{2} \alpha}{\beta} \right)^{2 (\ell - k)+1} \Gamma\left(\ell-k+ \frac{1}{2} \right) \ .
\end{eqnarray}
The coefficients $a_n$ are now given by Eq.~(\ref{eq555})

For comparison with previous work, we use our time-dependent wave function to calculate
$\langle x^2\rangle$, given by
\begin{eqnarray}
\langle x^2 \rangle &=&  \int \Psi^*(x,t)  \ x^2 \ \Psi(x,t) \ dx \ .
\end{eqnarray}
In terms of the $\psi_n(x)$ this is
\begin{eqnarray}
\langle x^2 \rangle &=&  \sum_{n , \ell}  \ a_n^* \ a_\ell \ e^{- i \omega_{n\ell} t} \ \int \psi_n(x) \ x^2 \ \psi_\ell(x) \ dx
\end{eqnarray}
where $\omega_{n\ell} \equiv E_n - E_\ell$. Using Eq.~(\ref{eq5})
this becomes
\beq
\langle x^2\rangle =  \sum_{n , \ell}  \ a_n^* \ a_\ell
e^{- i \omega_{n\ell} t}\sum_{k,j} d_{nk} d_{\ell j} (x^2)_{kj} \ .
\eeq
The result for $\langle x^2
\rangle^{1/2}$ is plotted in Fig.~\ref{FIG3}. As can be seen, the
result is vastly superior to Hartree-Fock, and on this scale
cannot be distinguished from that obtained using Fourier transform methods. The ratio of different
orders is shown in Fig.~\ref{FIG4}.
\begin{figure}
\begin{center}
\includegraphics[width=9cm]{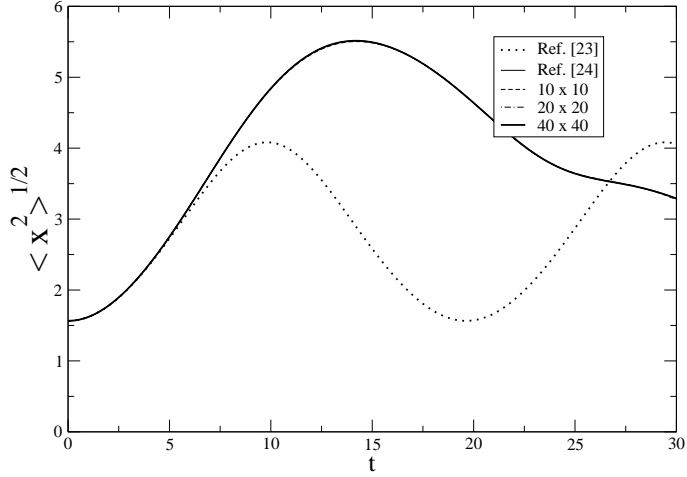}
\caption{$\langle x^2\rangle^{1/2}$ versus $t$
for the slow roll potential of Eq.~(\ref{dwell}).
Various orders of our method compared with the Hartree-Fock
method used in \cite{Cooper:1986wv} and the Fourier  transform method
of \cite{Lombardo:1999du}.}
\label{FIG3}
\end{center}
\end{figure}
\begin{figure}
\begin{center}
\includegraphics[width=9cm]{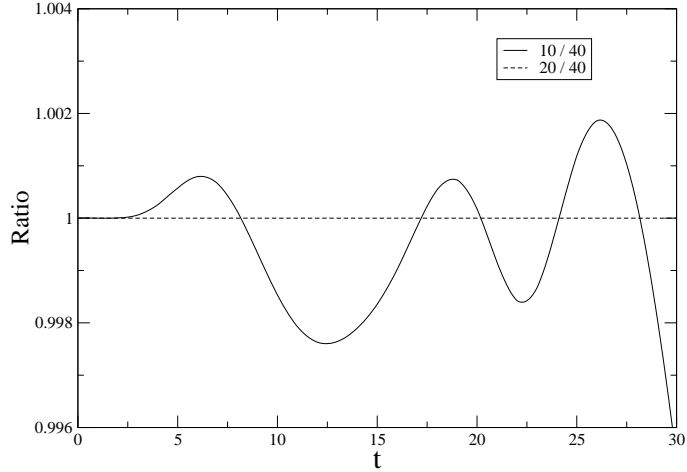}
\caption{Ratio of the results of our method for the slow roll potential:
orders $10$ and $20$ versus $40$. }
\label{FIG4}
\end{center}
\end{figure}

\subsubsection{Shifted Gaussian}\hfill\\

In this case we consider an initial wave function
$$\Psi(x,t=0) = \left(\frac{\mu }{2\pi} \right)^{1/4}
e^{- \mu  (x-x_o)^2/4},
$$
representing a particle localized around $x=x_o$ at $t=0$. As before, we obtain the coefficients $c_n$
by orthonormality:
\begin{eqnarray}
c_n &=& \int \phi_n^*(x) \Psi(x,t=0) dx \nonumber \\
&=& N_n \ \left(\frac{\mu }{2\pi} \right)^{1/4} \
\int_{-\infty}^{+\infty}
e^{- \frac{\beta^2}{2}(x-\frac{\mu x_o}{2\beta^2})^2}e^{-\mu x_o^2\alpha^2/4\beta^2}
H_n(\alpha x) dx ,
\end{eqnarray}
where now $\beta^2 \equiv \mu /2 + \alpha^2$. With a change of variable to
$y = \beta(x-\mu x_o/2\beta^2) $,
\begin{eqnarray}
c_n = \frac{N_n}{\beta} \left(\frac{\mu }{2\pi} \right)^{1/4}
e^{-\mu x_o^2\alpha^2/4\beta^2}
\int_{-\infty}^{+\infty}   e^{-y^2/2} H_n\left(\frac{\alpha}{\beta}\Big(y+
\frac{\mu x_o}{2\beta}\Big)\right) dy \ .
\end{eqnarray}
We expand the Hermite polynomial, to obtain
\begin{eqnarray} \nonumber
\hspace{-1cm}H_n(\alpha x) &=&
\sum_{k=0}^{[n/2]} (-1)^k \frac{n!}{(n-2 k)! k!} 2^{n-2 k}
\left(\frac{\alpha}{\beta}\right)^{(n-2k)}\left(y+\frac{\mu x_o}{2\beta}\right)^{n -2 k} \\ \cr
&=& \sum_{k=0}^{[n/2]} (-1)^k \frac{n!}{(n-2 k)! k!} 2^{n-2 k} \
\left(\frac{\alpha}{\beta}\right)^{(n-2k)}\ \sum_{j=0}^{n-2k} {n-2k \choose j} y^j\left(\frac{\mu x_o}{2\beta}\right)^{n-2k-j}
\nonumber \ .
\end{eqnarray}
Therefore 
\begin{eqnarray}
c_n &=& \frac{N_n}{\beta} \left(\frac{\mu }{2\pi} \right)^{1/4} e^{-\mu x_o^2\alpha^2/4\beta^2} \nonumber \\
&\times&\sum_{k=0}^{[n/2]}\sum_{j=0}^{n-2k} (-1)^k \frac{n! \ 2^{n-2 k}}{k! \ j! \ (n-2k-j)!}
\left(\frac{\alpha}{\beta}\right)^{n-2k}\left(\frac{\mu x_o}{2\beta}\right)^{n-2k-j} K_j \, ,
\end{eqnarray}
where
\begin{eqnarray}
K_j &\equiv& \int_{-\infty}^{+\infty}   y^{j}  e^{-y^2/2} dy = 2^{(j+1)/2} \ \left[ 1+ (-1)^j \right] \
\Gamma(j+ 1/2) \ .
\end{eqnarray}
Finally we obtain
\begin{eqnarray}\nonumber
c_{n} &=& \frac{N_{n}}{\beta}  \left(\frac{\mu }{2\pi} \right)^{1/4}
e^{-\mu x_o^2\alpha^2/4\beta^2}
\sum_{k=0}^{[n/2]} \sum_{j=0}^{[n/2]-k}
(-1)^k \frac{(n)! \ 2^{n-2 k}}{k! \ (2j)! \ (n-2k-2j)!} \\
& & \times \left(\frac{\alpha}{\beta}\right)^{n-2k}\left(\frac{\mu x_o}{2\beta}\right)^{n-2k-2j}
\ \left(\sqrt{2}\right)^{2j+1} \ \Gamma\left(j+1/2 \right) \ .
\end{eqnarray}

Figure~\ref{xaverage} shows the expectation value $\langle x \rangle$ as a function of time for four
different values of $\mu $. As expected, as $\mu $ increases, the frequency of oscillation between the
two wells increases as well. The plots in Fig.~\ref{xaverage} correspond to the values $g=1/2400$, $m=1/(2\sqrt{6})$
and $x_o = 5$.
\begin{figure}[ht]
\begin{center}
\includegraphics[width=9cm]{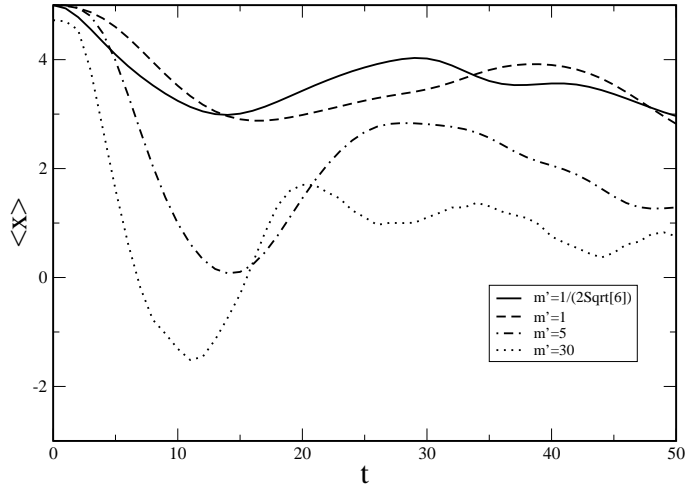}
\end{center}
\caption{$\langle x \rangle$ with four different choices of $\mu $
and with $x_o = 5$ for the slow roll potential with a shifted Gaussian.}
\label{xaverage}
\end{figure}

\section{Conclusions}
\label{concl}

We have shown that the matrix method, combined with the principle
of minimal sensitivity, is a powerful tool for finding the
spectrum of arbitrary polynomial potentials having only bound states.
The original method proposed in \cite{QM85} has been generalized in
two ways: for the calculation of higher levels, the matrix can be
centred around a higher levels of the SHO, and for noneven potentials 
the introduction of a shift parameter
improves the accuracy of the method for a given dimension $N$. This
aspect of the method could be used for a general, non-polynomial potential,
in conjunction with its Taylor expansion about a given point.

As a by-product of the calculation of the spectrum, the method also
provides approximations to the energy eigenfunctions. Knowing both
the energies and their eigenfunctions we are then able to implement
the method of stationary states to track the time development of a
given initial wave function with good accuracy and for long time
scales, as we have demonstrated for the slow-roll potential.

\bigskip

P.A. acknowledges support of Conacyt grant no. C01-40633/A-1 and of Alvarez-Buylla
fund of the University of Colima.
A.A. acknowledges support from Conacyt grant no. 44950 and PROMEP.


\newpage
\section*{References}
\begin{thebibliography}{}

\bibitem{DM} A.~M. Duncan and M. Moshe, {\it Phys. Lett.} {\bf B215}, 352 (1988).
\bibitem{HFJ}{A. Duncan and H.~F.Jones,
{\it Phys. Rev.} {\bf D47}, 2560 (1993).}
\bibitem{AFC90}  G.~A. Arteca, F.~M. Fern\'{a}ndez, and E.~A. Castro, ``Large
order perturbation theory and summation methods in quantum mechanics''
(Springer, Berlin, Heidelberg, New York, London, Paris, Tokyo, Hong Kong,
Barcelona, 1990).
\bibitem{Jan95} W. Janke and H. Kleinert, {\it Phys. Rev. Lett.} {\bf 75}, 2787 (1995).
\bibitem{Yuk91}  V.~I. Yukalov,{\it J. Math. Phys.} {\bf 32}, 1235 (1991).
\bibitem{AAD:04} P. Amore, A. Aranda and A. De Pace, {\it Journal of Physics} {\bf A37}, 3515 (2004).
\bibitem{AADL:04} P. Amore, A. Aranda,A. De Pace, and J. L\'opez, {\it Phys. Lett.} {\bf A329}, 451 (2004).
\bibitem{PMS} P.~M. Stevenson, {\it Phys. Rev.} {\bf D23}, 2916 (1981).
\bibitem{AA1:03} P. Amore and A. Aranda, {\it Phys. Lett.} {\bf A316}, 218 (2003).
\bibitem{AA2:03} P. Amore and A. Aranda, \textit{Accepted for publication in Journal of Sound and Vibration}
[arXiv:math-ph/0303052].
\bibitem{AM:04} P. Amore and H. Montes, {\it Phys. Lett.} {\bf B327} 158 (2004) .
\bibitem{AS:04} P. Amore and R. S\'{a}enz, [arXiv:math-ph/0405030].
\bibitem{AAFS:04} P. Amore, A.  Aranda, F. Fern\'andez and R. S\'aenz, [arXiv:math-ph/0407014].
\bibitem{AF:04} P. Amore and F. Fern\'andez, [arXiv:math-ph/0409034].
\bibitem{AL:04}  P. Amore and J. L\'opez, [arXiv:quant-ph/0405090].
\bibitem{Amore:04} P. Amore, [arXiv:math-ph/0408036].
\bibitem{Jones:2000au} H.~F. Jones, P. Parkin and D. Winder,
{\it Phys. Rev.} {\bf D63}, 125013 (2001) [arXiv:hep-th/0008069].
\bibitem{QM85} R.~M. Quick and H.~G. Miller,  {\it Phys. Rev.} {\bf D31}, 2682 (1985).
\bibitem{Morales} J. Morales, J. L\'opez-Vega and A. Palma, {\it J. Math. Phys.} {\bf 28}, 1032 (1987).
\bibitem{mei97} H. Meissner and O. Steinborn, {\it Phys. Rev.} {\bf A56}, 1189 (1997).
\bibitem{Maluendes1982}
S.~A. Maluendes, G.~A. Arteca, F.~M. Fern\'andez and E.~A. Castro, {\it Mol. Phys.} {\bf 45}, 511 (1982)
\bibitem{Guth:1985ya} A.~H. Guth and S.~Y. Pi, {\it Phys. Rev.} {\bf D32}, 1899 (1985).
\bibitem{Cooper:1986wv} F. Cooper, S.~Y. Pi and P.~N. Stancioff, {\it Phys. Rev.} {\bf D34}, 3831 (1986).
\bibitem{Lombardo:1999du}F.~C. Lombardo, F.~D. Mazzitelli and D. Monteoliva, {\it Phys. Rev.} {\bf D62}, 045016 (2000) [arXiv:hep-ph/9912448].

\verb''\end{thebibliography}
\end{document}